\documentclass[12pt]{article}
\usepackage{amsmath}
\usepackage{amssymb}
\usepackage{amsfonts}
\usepackage{amscd}
\usepackage{amsbsy}
\usepackage{amsthm}
\usepackage{latexsym}
\usepackage{graphicx}
\usepackage{slashed}
\usepackage{color}
\def\nn{\nonumber}       
\def\n{\label}                 
\def\r{\ref}                    

\newcommand{\eq}[1]{(\ref{#1})}

\def\beq{\begin{eqnarray}}
\def\eeq{\end{eqnarray}}

\def\al{\alpha}

\def\de{\delta}
\def\vp{\varepsilon}

\def\la{\lambda}

\def\si{\sigma}
\def\om{\omega}

\def\Ga{\Gamma}

\def\Om{\Omega}

\usepackage{indentfirst}

\begin{document}

\begin{center}

{\Large \bf Functional renormalization group approach and gauge
dependence in gravity theories}

\vskip 8mm {\bf V\'{\i}tor F. Barra}$^{a}$\footnote{E-mail address:
\ vitorbarra@ice.ufjf.br}, \  \ {\bf Peter M.
Lavrov}$^{b,c}$\footnote{E-mail address: \ lavrov@tspu.edu.ru}, \ \
{\bf Eduardo Antonio dos Reis}$^{a}$\footnote{E-mail address: \
eareis@fisica.ufjf.br},
\\
{\bf Tib\' erio de Paula Netto}$^{d}$\footnote{
E-mail address: \ tiberio@sustech.edu.cn}, \ \
{\bf Ilya L. Shapiro}$^{a,b,c}$\footnote{E-mail address: \
shapiro@fisica.ufjf.br} \vskip 4mm

$^a$ \
{\sl
Departamento de F\'{\i}sica, \ ICE, \ Universidade Federal de Juiz de Fora,
36036-330 Juiz de Fora, \ MG, \ Brazil}
\vskip 2mm

$^b$ \
{\sl
Department of Theoretical Physics, Tomsk State Pedagogical
University,  634061 Tomsk, Russia}
\vskip 2mm

$^c$ \
{\sl
National Research Tomsk State University, 634050 Tomsk, Russia}
\vskip 2mm

$^d$ \
{\sl
Departament of Physics, Southern University of Science and Technology,
518055 Shenzhen, China}
\end{center}

\centerline{{\bf Abstract}}

\begin{quotation}
\noindent
We investigate the gauge symmetry and gauge fixing dependence
properties of the effective average action for quantum gravity
models of general form. Using the background field formalism and
the standard BRST-based arguments, one can establish the special
class of regulator functions that preserves the background field
symmetry of the effective average action. Unfortunately, regardless
the gauge symmetry is preserved at the quantum level, the
non-invariance of the regulator action under the global BRST
transformations leads to the gauge fixing dependence even under
the use of the on-shell conditions.
\vskip 4mm

\noindent
{\bf Keywords:} quantum gravity, background field method,
functional renormalization group approach, gauge dependence

\vskip 4mm

\end{quotation}

\section{Introduction}
\label{int}

The interest to the non-perturbative formulation in quantum gravity
has two strong motivations. First, there are a long-standing
expectations that even the perturbatively non-renormalizable models
such as the simplest quantum gravity based on general relativity may
be quantum mechanically consistent due to the asymptotic safety
scenario \cite{Weinberg-79} (see \cite{NiederReuter,Percacci-2007}
for comprehensive reviews). On the other hand, there is a possibility
that the non-perturbative effects may provide unitarity in the fourth
derivative theory by transforming the massive unphysical pole, which
spoils the spectrum of this renormalizable theory \cite{Stelle77}.
The presence of such a massive ghost violates the stability of
classical solutions (see e.g. discussion and further references in
\cite{HDQG}). At the quantum level, the perturbative information
is insufficient to conclude whether in the dressed propagator the ghost
pole does transform into a non-offensive pair of complex conjugate
poles \cite{salstr}.

The perturbative renormalization group in this model is well-explored
\cite{frts82,avbar86,Gauss}, but it is not conclusive for the discussion
of the dressed propagator. In general, the attempts to explore this
possibility in the semiclassical and perturbative quantum level
\cite{Tomboulis-77,antomb} has been proved non-conclusive
\cite{Johnston}, and hence the main hope is
related to the non-perturbative calculations in the framework of
Functional Renormalization Group approach \cite{CodPer-06}
(see \cite{CodPerRah-2009} for an extensive review).

Thus, in both cases the consistency of the results obtained within
Functional Renormalization Group approach is of the utmost
importance. In this respect there are two main dangers. For the
quantum gravity models based on general relativity, the running of
Newton constant in four-dimensional spacetime is always obtained on
the basis of quadratic divergences. These divergences are known to
have strong regularization dependence. In particular, they are
absent in dimensional regularization and can be freely modified in
all known cut-off schemes by changing the regularization parameter.
This part of the problem does not exist for the Functional
Renormalization Group applied to the fourth derivative quantum
gravity. However, in this case there is yet another serious problem,
related to the gauge-fixing dependence of the effective average
action. This problem is the subject of the present work. In the
previous publications \cite{Lavrov:2012xz,Lav(in),FRG-ELT} we
explored the gauge fixing dependence in Yang-Mills theories and it
was shown that such dependence for the effective average action does
not vanish on-shell, except
maybe in the fixed point where
this object becomes identical with the usual effective action in
quantum field theory. Except this special point there is
uncontrollable dependence on the set of arbitrary gauge fixing
parameters, and thus one can expect a strong arbitrariness in the
renormalization group flows which lead to the fixed point and, in
fact, define its position and proper existence. The main purpose of
the present work is to extend this conclusion to quantum gravity. It
is remarkable that one can complete this task for the quantum
gravity theory of an arbitrary form, without using the concrete form
of the action. One can use this consideration, e.g. for the
superrenormalizable models of quantum gravity, when the perturbative
renormalization group may be exact and, moreover, completely
independent on the gauge fixing \cite{highderi,QG-betas}. This
example is somehow the most explicit one, since it shows that the
transition from standard quantum field theory to the functional
renormalization group (FRG) may actually spoil the ``perfect''
situation, namely exact and universal renormalization group flow.

In Yang-Mills theories \cite{YMtheories}, the gauge symmetry of
the initial action is broken on quantum level due to the gauge fixing
procedure in the process of quantization. In turn, the effective
potential depends on gauge \cite{Jackiw,Dolan,Nielsen,Fukuda}.
This dependence occur in a special way, such that it disappears on-shell
\cite{LavTyu,Voronov}, which means that it is possible to give
physical interpretation to the results obtained at the quantum level.

One of the well-developed non-perturbative methods in Quantum Field
Theory to study quantum properties of physical models beyond the
perturbation theory is the FRG approach \cite{Wetterich,Wetterich2}
(see also the review papers
\cite{Berges,Bagnuls,Polonyi,Pawlowski,Delamotte,Rosten}). As we
have already mentioned above, when applied to gauge theories, this
method leads to obstacles related to the  on-shell gauge dependence
of the  effective average  action.

There are some efforts to solve this problems. One of them consists
from reformulation of Yang-Mills theory with the application of a
gauge-invariant cut-off dependent regulator function that is
introduced as a covariant form factor into the action of Yang-Mills
fields, which leads to an invariant regulator action
\cite{Morris,Arnone}. As a consequence, the
effective average action is gauge invariant on-shell. A second approach
\cite{Branchina, Pawlowski2} is based in the use of the
Vilkovisky-DeWitt covariant effective action
\cite{Vilkovisky,Vilkovisky2}. This technique provides gauge
independence even off-shell, but it introduce other types of
ambiguities. An alternative formulation was presented in
\cite{Lavrov:2012xz}, it consists of an alternative way of
introducing the regulator function as a composite operator. When
applied in gauge theories, this approach leads to the  on-shell gauge
invariance of the effective average  action.

In the present work, we apply the background field method
\cite{DeWitt-BFM, Arefeva, Abbott} (recent advances for the gauge
theories can be found in \cite{Barvinsky, Batalin-Lavrov,
Frenkel,Lav(in),Batalin-Lavrov-Tyutin,LavShap,Breno} and discussion
for the quantum gravities case in \cite{Percacci}) in the FRG
approach as a reformulation to the quantization procedure for
quantum gravity theories to study the gauge dependence problem in
this kind of theory. This method allows us to work with the
effective action which is invariant under the gauge transformations
of the background fields.

Despite the numerous aspects of quantum properties successfully
studied with the background field method,
\cite{Hooft, Kluberg, Grisaru, Capper, Ichinose, Goroff, Ven,
Grassi, Becchi, Ferrari}, the gauge dependence problem remains
important \cite{Lav(in)} and need to be considered in more details.
We obtain restrictions on the tensor structure of the regulator functions
which allows us to construct a gauge invariant effective average
action. Nevertheless, the effective average  action remains dependent
of the gauge choice at on-shell level.

The paper is organized as follows. In section \r{sect2} we introduce
general considerations of quantum gravity theories through the
background field method. In section \r{sect3} we consider the FRG
approach for quantum gravity theories and find conditions that we
must impose in the regulator functions to maintain the gauge invariance
of background effective average  action. Based on this, we also present
some interesting candidates to the regulator functions. In section
\r{sect4} we prove the gauge dependence of vacuum functional for
the model under consideration. Finally, our conclusions and remarks
are presented in section \r{sect5}.

In the paper, DeWitt notations \cite{DeWitt} are used. The short
notation for integration in $D$ dimension is $\int d^Dx=\int dx$.
All the derivatives with respect to fields are left derivative
unless otherwise stated. The Grassmann parity of a quantity
$A$ is denoted as $\vp(A)$.

\section{Quantum gravity in the background field formalism}
\label{sect2}

Let us consider an arbitrary action $S_0 = S_0(g)$, where
$g=g_{\alpha\beta}(x)$ is the metric
tensor of an arbitrary Riemann manifolds. We assume that the
action is invariant under general coordinates transformations
\beq
x^\mu \to x^\mu = x^\mu (x^\prime),
\eeq
which leads to the metric transformations
\beq
g_{\alpha\beta} ^\prime (x^\prime)
= g_{\mu \nu}(x) \frac{\partial x^\mu}{\partial x^{\prime\alpha}}
\frac{\partial x^\nu}{\partial x^{\prime\beta}}
\eeq
and consider the infinitesimal form of these transformations,
$x'^\sigma = x^\sigma + \xi^\sigma (x)$, when
\beq
\delta g_{\alpha\beta}
\,=\,
-\xi^\sigma (x)\partial_\sigma g_{\alpha\beta}(x)
-g_{\alpha\sigma}(x)\partial_\beta\xi^\sigma(x)
-g_{\sigma\beta}(x)\partial_\alpha\xi^\sigma(x).
\n{eq:ten-trans}
\eeq
The diffeomorphism (\r{eq:ten-trans}) can be considered as
the gauge transformation for
$g_{\alpha\beta}(x)$ 
\beq
\delta g_{\alpha\beta}(x)
= \int dy\, R_{\alpha\beta\sigma}(x,y;g)\,\xi^\sigma(y),
\eeq
where
\beq
R_{\alpha\beta\sigma}(x,y;g)= -\delta(x-y)\partial_\sigma g_{\alpha\beta}(x)
-g_{\alpha\sigma}(x)\partial_\beta\delta(x-y)
-g_{\sigma\beta}(x)\partial_\alpha\delta(x-y)
\eeq
are the generators of the gauge transformations of the metric tensor,
and $\xi^\sigma(y)$ are the gauge parameters. The algebra of  generators
is closed, namely
\beq
\int du \bigg[
\frac{\delta R_{\alpha\beta\sigma}(x,y;g)}{\delta g_{\mu\nu}(u)}
R_{\mu\nu\gamma}(u,z;g)
-
\frac{\delta R_{\alpha\beta\gamma}(x,z;g)}{\delta g_{\mu\nu}(u)}
R_{\mu\nu\sigma}(u,y;g)
\bigg]
\nn
\\
=
-\int du R_{\alpha\beta\lambda}(x,u;g)F^\lambda _{\sigma\gamma}(u,y,z),
\n{eq:gauge-alg}
\eeq
where
\begin{align}
F^\lambda _{\mu\nu}(x,y,z)&= \delta(x-y)
\delta^\lambda _\nu\partial_\mu\delta(x-z)
-\delta(x-z)\delta^\lambda _\mu \partial_\nu \delta(x-y),
\\
\mbox{with}\quad
F^\lambda _{\mu\nu}(x,y,z)&=-F^\lambda _{\nu\mu}(x,z,y)
\n{eq:stru-func}
\end{align}
are structure functions of the algebra which does not depend on the
metric tensor. Let us stress that the mentioned features are valid for
an arbitrary action of gravity, since the algebra presented above is
independent on the initial action. Therefore, any theory of gravity is
a gauge theory and the structure functions are independent of the
fields, that means quantum gravity is similar to the Yang-Mills theory.

An useful procedure to quantize gauge theories is the so called
background field formalism. In what follows, we shall perform the
quantization of gravity on an arbitrary external background metric
$\bar{g}_{\alpha\beta}(x)$. The standard references on the
background field formalism in quantum field theory are
\cite{DeWitt-BFM, Arefeva, Abbott} (see also recent advances for the
gauge theories in \cite{Lav(in)} and \cite{LavShap} for the
discussion of quantum gravity).

In the background field formalism, the metric tensor
$g_{\alpha\beta}(x)$ is replaced by the sum
\beq
g_{\alpha\beta} (x) =\bar{g}_{\alpha \beta}(x) + h_{\alpha \beta}(x),
\eeq
where $\bar{g}_{\alpha \beta}(x)$ is an external (background) metric
field and $h_{\alpha\beta}(x)$ is the variable of integration, also called
quantum metric. Thus, the initial action is replaced by
\beq
S_0 (g) \to S_0 (\bar{g}+h).
\nn
\eeq

The Faddeev-Popov $S_{FP}(\phi ,\bar{g})$ action is constructed
in the standard way \cite{FPaction}
\beq
S_{FP}(\phi ,\bar{g}) =
S_0 (\bar{g}+h) + S_{gh}(\phi ,\bar{g}) + S_{gf}(\phi ,\bar{g}),
\n{eq:Sfp}
\eeq
where $S_{gh}(\phi ,\bar{g})$ is the ghost action and $S_{gf}(\phi ,\bar{g})$
is the gauge fixing action. In the presence of external metric
$\bar{g}_{\alpha\beta}(x)$, they can be written as
\beq
S_{gh}(\phi ,\bar{g})
=
\int dxdydz \sqrt{-\bar{g}(x)}
\bar{C}^\alpha(x)H^{\beta\gamma} _\alpha(x,y;\bar{g},h)
R_{\beta\gamma\sigma}(y,z;\bar{g}+h)C^\sigma(z),
\eeq
\beq
S_{gf}(\phi ,\bar{g}) =
\int dx \sqrt{-\bar{g}(x)}
B^\alpha(x) \chi_\alpha (x;\bar{g},h),
\eeq
where
\beq
H^{\beta\gamma} _\alpha(x,y;\bar{g},h)
=
\frac{\delta \chi_\alpha(x;\bar{g},h)}{\delta h_{\beta\gamma}(y)}
\eeq
and $\chi_\alpha(x;\bar{g},h)$ are the gauge fixing functions,
$\phi^i(x) = \{h_{\alpha\beta}(x),B^\alpha(x), C^\alpha(x) ,
\bar{C}^\alpha(x)\}$
is the set of quantum fields, $C^\alpha(x)$ and $\bar{C}^\alpha(x)$
are the ghost and anti-ghost fields, respectively, and $B^\alpha (x)$
are the Nakanishi-Lautrup auxiliary fields. The Grassmann parity of
all quantum fields are as follows,
\beq
\varepsilon(h_{\alpha\beta}) = \varepsilon(B^\alpha)=0,
\qquad
\varepsilon(\bar{C}^\alpha) = \varepsilon(C^\alpha)=1,
\qquad
\varepsilon(\phi^i) = \varepsilon_i.
\nn
\eeq
The ghost numbers are
\beq
\mathrm{gh}(B^\alpha)=\mathrm{gh}(h_{\alpha\beta})=0,
\qquad
\mathrm{gh}(C^\alpha) = 1,
\qquad
\qquad
\mathrm{gh}(\bar{C}^\alpha) = -1.
\nn
\eeq

Independently of gauge fixing function choice, the action
(\ref{eq:Sfp}) is invariant under a global supersymmetry
transformation, known as BRST symmetry \cite{BRS,BRST}. The
gravitational BRST transformations were introduced in
\cite{Delbourgo, Stelle77, Townsend} and can be presented as
\begin{align}
\delta_B h_{\alpha\beta} (x)
&=
-\left(C^\sigma(x)\partial_\sigma g_{\alpha\beta}(x)
+
g_{\alpha\sigma}(x)\partial_\beta C^\sigma(x)
+
g_{\sigma\beta}(x)\partial_\alpha  C^\sigma(x)\right)\lambda ,
\nn
\\
\delta_B C^\alpha (x)
&=
C^\sigma (x) \partial_\sigma C^\alpha(x) \lambda ,
\nn
\\
\delta_B \bar{C}^\alpha (x)
&=
B^\alpha (x) \lambda  ,
\nn
\\
\delta_B B^\alpha (x)
&=
0,
\n{eq:BRST}
\end{align}
where $\lambda$ is a constant Grassmann parameter.  In condensed
notation, we can write the BRST transformations as \beq \delta_B
\phi^i(x) = R^i(x;\phi ,\bar{g})\lambda, \qquad \varepsilon(R^i) =
\varepsilon_i +1, \n{eq:BRSTcon} \eeq where $R^i = \{R_{\alpha\beta}
^{(h)} ,R_{(B)} ^\alpha ,R_{(C)} ^\alpha ,R_{(\bar{C})} ^\alpha \}$
and
\begin{align}
R_{\alpha\beta} ^{(h)} (x;\phi ,\bar{g})
&=
-C^\sigma(x)\partial_\sigma g_{\alpha\beta}(x)
-
g_{\alpha\sigma}(x)\partial_\beta C^\sigma(x)
-
g_{\sigma\beta}(x)\partial_\alpha C^\sigma(x) ,
\nn
\\
R_{(C)} ^\alpha (x;\phi ,\bar{g})
&=
C^\sigma (x) \partial_\sigma C^\alpha(x),
\nn
\\
R_{(\bar{C})} ^\alpha (x;\phi ,\bar{g})
&=
B^\alpha (x),
\nn
\\
R_{(B)} ^\alpha(x;\phi ,\bar{g})
&=
0.
\end{align}

The generating functional of Green functions is defined as \beq Z(J,
\bar{g}) = \int d\phi \exp\Big\{\frac{i}{\hbar}[S_{FP}(\phi
,\bar{g})+J\phi]\Big\} =
\exp\Big\{\frac{i}{\hbar}W(J,\bar{g})\Big\}, \n{eq:functional} \eeq
where $W(J,\bar{g})$ is the generating functional of connected Green
functions. In Eq.~\eq{eq:functional} the product of the sources
$J_i(x)$   and the fields $\phi^i(x)$ was written in the condensed
notation of DeWitt \cite{DeWitt}.
Explicitly,
\beq
J\phi = \int dx J_i (x) \phi^i (x),
\qquad
\mathrm{where}
\qquad
J_i(x) = \{J^{\mu\nu}, J^{(B)} _\alpha(x), \bar{J}_\alpha(x), J_\alpha(x)\}
\eeq
with the Grassmann parities $\varepsilon(J_i)=\varepsilon(\phi^i)$
and ghost numbers $\mathrm{gh}(J_i)=\mathrm{gh}(\phi^i)$.

The effective action $\Gamma(\Phi ,\bar{g})$ is defined in terms of
the Legendre transformation \beq \Gamma(\Phi ,\bar{g}) =
W(J,\bar{g}) - J \Phi , \eeq
where $\Phi = \{{\Phi^i}\}$ are the
mean fields and the $J_i$  are the solution of the equation \beq
\frac{\delta W(J,\bar{g})}{\delta J_i} = \Phi^i .
\eeq

It is well-known \cite{LavTyu,Voronov} that the effective action is
gauge independent on-shell, \beq \delta \Gamma(\Phi ,\bar{g})\Big|
_{\frac{\delta\Gamma(\Phi , \bar{g})}{\delta\Phi}=0}=0. \eeq

At this moment we have considered only the transformations of the
quantum fields. However, the background metric also transform
together with the quantum fields  in the so-called background field
transformations. The rules of such transformation can be written, in
the local formulation, as
\begin{align}
\delta_\omega \bar{g}_{\alpha\beta}(x)
&=
-\partial_\sigma \bar{g}_{\alpha\beta}(x)\omega^\sigma
-\bar{g}_{\alpha\sigma}(x)\partial_\beta\omega^\sigma
-\bar{g}_{\sigma \beta}(x)\partial_\alpha\omega^\sigma ,
\nn
\\
\delta _{\omega} h_{\alpha\beta}(x)
&=
-\partial_\sigma h_{\alpha\beta}(x)\omega^\sigma
-h_{\alpha\sigma}(x) \partial_\beta\omega^\sigma
-h_{\sigma\beta}(x) \partial_\alpha\omega^\sigma ,
\nn
\\
\delta_{\omega}\bar{C}^\alpha (x)
&=
-\omega^\sigma\partial_\sigma\bar{C}^\alpha(x)
+\bar{C}^\sigma(x)\partial_\sigma\omega^\alpha ,
\nn
\\
\delta _{\omega}C^\alpha(x)
&=
-\omega^\sigma\partial_\sigma C^\alpha(x)
+C^\sigma (x)\partial_\sigma\omega^\alpha ,
\nn
\\
\delta_{\omega}B^\alpha (x)
&=
-\omega^\sigma\partial_\sigma B^\alpha(x)
+B^\sigma(x)\partial_\sigma\omega^\alpha
,
\n{eq:backgauge}
\end{align}
where $\omega^\sigma = \omega^\sigma (x)$ are arbitrary functions.
The background field transformations have the same structure of
tensor transformations for tensors of types $(0,2)$ and $(1,0)$. The
background invariance of Faddeev-Popov action for quantum gravity
is known \cite{LavShap} and reads 
\beq
\delta_\omega S_{FP}(\phi , \bar{g}) = 0
.
\n{eq:FPinva}
\eeq

A consequence of (\r{eq:FPinva}) is the gauge invariance of
(\r{eq:functional}).  Namely, \beq \delta_\omega Z(J,\bar{g})=0.
\eeq From this expression it is possible to prove that
$\Gamma(\Phi,\bar{g})$
is also gauge invariant
\beq
\delta_\omega \Gamma(\Phi,\bar{g})=0.
\eeq

In the next sections we will discuss this and other features  for
quantum gravity theories in the framework of the FRG approach.

\section{FRG approach for quantum gravity theories}
\label{sect3}

The main idea of functional renormalization group (FRG) is to use
instead of $\Gamma$ an effective average  action, $\Gamma_k$, where
$k$ is a momentum-shell parameter \cite{Wetterich}, in a way that
\beq
\lim_{k\to 0} \Gamma_k (\phi,\bar{g}) = \Gamma(\phi,\bar{g}) .
\eeq

In order to obtain $\Gamma_k (\phi,\bar{g})$, we introduce the average action
\beq
S_{k FP}(\phi ,\bar{g}) = S_{FP}(\phi ,\bar{g})+S_k(\phi ,\bar{g}),
\eeq
where $S_k(\phi , \bar{g})$ is the scale-dependent regulator action defined
in curved spacetime
\beq
S_k(\phi , \bar{g})
=
\int dx \sqrt{-\bar{g}(x)} \mathcal{L} _k(\phi, \bar{g})
\n{eq:act}
\eeq
and the Lagrangian density is written as
\beq
\mathcal{L} _k(\phi , \bar{g})
=
\frac{1}{2}h_{\alpha \beta}(x)
R^{(1) \alpha \beta \gamma \delta} _{k} (x; \bar{g})
h_{\gamma \delta}(x)
+
\bar{C}^{\alpha}(x) R^{(2)} _{k \alpha \beta} (x; \bar{g})C^{\beta} (x)
\n{eq:lagr}
,
\eeq
where the regulator functions
$R^{(1)\alpha \beta \gamma \delta} _{k} (x; \bar{g})$ and
$R^{(2)} _{k \alpha \beta} (x; \bar{g})$ are dependent on the external fields
$\bar{g} _{\alpha \beta}(x)$. The regulator functions obey the properties
\beq
\lim_{k\to 0}R^{(1)\alpha \beta \gamma \delta} _{k} (x; \bar{g}) = 0
\qquad
\mathrm{and}
\qquad
\lim_{k\to 0}R^{(2)} _{k \alpha \beta} (x; \bar{g}) =0,
\eeq
which means that the average action recovers the
Faddeev-Poppov action (\r{eq:Sfp}) in the limit
when $k\to 0$. The regulator functions
$R^{(1)\alpha \beta \gamma \delta} _{k} (x; \bar{g})$ also
obey, by construction, the symmetry properties
\beq
R^{(1)\alpha \beta \gamma \delta} _{k} (x; \bar{g})
=
R^{(1)\beta \alpha \gamma \delta} _{k} (x; \bar{g})
=
R^{(1) \alpha \beta \delta \gamma} _{k} (x; \bar{g})
=
R^{(1)\gamma \delta \alpha \beta } _{k} (x; \bar{g})
.
\n{eq:symmetry}
\eeq

We want to solve the problem of average action invariance under
background field transformations, namely \beq \delta_\omega
S_{kFP}(\phi ,\bar{g}) = \delta_\omega S_k(\phi,\bar{g})= 0
\n{eq:inva}, \eeq where the relation (\r{eq:FPinva}) is used.

In what follows, we present explicit calculation of variation of action
(\ref{eq:act}). For this purpose, we write (\ref{eq:inva}) as
\beq
\delta_\omega S_k(\phi , \bar{g})
=
\int dx  \Big\{
\delta_\omega \sqrt{-\bar{g}(x)}\mathcal{L} _k(\phi ,\bar{g})
+
\sqrt{-\bar{g}(x)}\delta_\omega \mathcal{L} _k(\phi ,\bar{g})
\Big\}
=0.
\n{eq:twoterms}
\eeq

For the first term in (\r{eq:twoterms}) we have
\begin{equation}
\begin{split}
\n{ds1}
\int dx
\delta_\omega \sqrt{-\bar{g}(x)}
\mathcal{L} _k(\phi , \bar{g})
&=
-\int dx\partial_\sigma(\sqrt{-\bar{g}(x)}\omega^\sigma)
\mathcal{L} _k(\phi , \bar{g}) 
\\
&=
\int dx\sqrt{-\bar{g}(x)}\omega^\sigma
\partial_\sigma \mathcal{L} _k(\phi, \bar{g}),
\end{split}
\end{equation}
where integration by parts was used. 

The variation of $\mathcal{L} _k(\phi , \bar{g})$ in second term
of equation (\r{eq:twoterms}) can be presented as
\begin{equation}
\begin{split}
\delta_\omega \mathcal{L} _k(\phi , \bar{g})
=& \,\,
\frac{1}{2}\delta_\omega h_{\alpha \beta} (x)
R^{(1)\alpha \beta \gamma \delta} _k (x;\bar{g})
h_{\gamma \delta}(x)
+
\frac{1}{2} h_{\alpha \beta} (x)
\delta_\omega  R^{(1)\alpha \beta \gamma \delta} _k (x;\bar{g})
h_{\gamma \delta}(x)
\\
&+
\frac{1}{2}h_{\alpha \beta} (x)
R^{(1)\alpha \beta \gamma \delta} _k (x;\bar{g})
\delta_\omega h_{\gamma \delta}(x)
+
\delta_\omega  \bar{C}^\alpha (x)
R^{(2)} _{k \alpha \beta} (x;\bar{g})
C^\beta (x)
\\
&+
\bar{C} ^\alpha (x)
\delta_\omega R^{(2)} _{k \alpha \beta} (x;\bar{g})
C^\beta (x)
+
\bar{C}^\alpha (x)
R^{(2)} _{k \alpha \beta} (x;\bar{g})
\delta_\omega C^\beta (x).
\end{split}
\end{equation}

In terms of transformations (\r{eq:backgauge}), the above expression reads
\begin{equation}
\begin{split}
\n{ds2}
\delta_\omega \mathcal{L} _k(\phi, \bar{g})
=&
-\frac{1}{2}
\big(\partial_\sigma h_{\alpha \beta}(x) \omega^\sigma
+h_{\alpha \sigma}(x) \partial_\beta \omega^\sigma
+h_{\sigma \beta}(x) \partial_\alpha \omega^\sigma
\big)
R^{(1)\alpha \beta \gamma \delta} _k (x;\bar{g})
h_{\gamma \delta}(x)
\\
&-
\frac{1}{2}h_{\alpha \beta}
R^{(1)\alpha \beta \gamma \delta} _k (x;\bar{g})
\big(
\partial_\sigma h_{\gamma \delta}(x) \omega^\sigma
+h_{\gamma \sigma}(x) \partial_\delta \omega^\sigma
+h_{\sigma \delta}(x) \partial_\gamma \omega^\sigma
\big)
\\
&+
\frac{1}{2} h_{\alpha \beta}(x)
\delta_\omega R^{(1)\alpha \beta \gamma \delta} _k (x;\bar{g})
h_{\gamma \delta}(x)
+
\bar{C}^\alpha (x)
\delta_\omega R^{(2)} _{k \alpha \beta} (x;\bar{g})
C^\beta (x)
\\
&+
\big(
-\omega^\sigma \partial_\sigma \bar{C}^\alpha (x)
+\bar{C}^\sigma (x) \partial_\sigma \omega ^\alpha
\big)
R^{(2)} _{k \alpha \beta} (x;\bar{g})
C^\beta (x)
\\
&+
\bar{C}^\alpha (x)
R^{(2)} _{k \alpha \beta} (x;\bar{g})
\big(-\omega^\sigma \partial_\sigma C^\beta (x)
+C^\sigma (x) \partial_\sigma \omega ^\beta\big).
\end{split}
\end{equation}
Thus, by means of Eqs.~\eq{ds1} and \eq{ds2} the variation of
the action can be written in the compact way
\beq
\delta_\omega S_k (\phi, \bar{g})
=
\int dx \sqrt{-\bar{g}(x)} \Big\{
\frac{1}{2}h_{\alpha \beta}(x)
M^{(1) \alpha \beta \gamma \delta} _{\omega k} (x;\bar{g})
h_{\gamma \delta} (x)
+
\bar{C}^\alpha (x)
M^{(2)} _{\omega k \alpha \beta} (x;\bar{g})
C^\beta (x)
\Big\}
,
\eeq
where
\begin{equation}
\begin{split}
M^{(1) \alpha \beta \gamma \delta} _{\omega k} (x;\bar{g})
=& \,\,
\delta _\omega  R^{(1)\alpha \beta \gamma \delta} _k (x;\bar{g})
+
\omega ^\sigma
\partial_\sigma R^{(1)\alpha \beta \gamma \delta} _k (x;\bar{g})
-
\partial_\sigma \omega ^\alpha
R^{(1)\sigma \beta \gamma \delta} _k (x;\bar{g})
\nn
\\
&-
\partial_\sigma \omega ^\beta
R^{(1)\alpha \sigma \gamma \delta} _k (x;\bar{g})
-
\partial_\sigma \omega ^\gamma
R^{(1)\alpha \beta \sigma \delta} _k (x;\bar{g})
- \partial_\sigma \omega ^\delta
R^{(1)\alpha \beta \gamma \sigma} _k (x;\bar{g})
\n{eq:M1}
\end{split}
\end{equation}
and
\beq
M^{(2)} _{\omega k \alpha \beta} (x;\bar{g})
=
\delta_\omega R^{(2)} _{k \alpha \beta} (x;\bar{g})
+
\omega ^\sigma
\partial_\sigma R^{(2)} _{k \alpha \beta} (x;\bar{g})
+
R^{(2)} _{k \sigma \beta} (x;\bar{g})
\partial_\alpha \omega ^\sigma
+
R^{(2)} _{k \alpha \sigma} (x;\bar{g})
\partial_\beta \omega^\sigma
.
\n{eq:M2}
\eeq

In order to ensure the invariance of (\r{eq:act}), it is
necessary that the following conditions are satisfied
\beq
M^{(1) \alpha \beta \gamma \delta} _{\omega k} (x;\bar{g})
=
0
\qquad
\mathrm{and}
\qquad
M^{(2)} _{\omega k \alpha \beta} (x;\bar{g})
=
0.
\eeq
As a result, we obtain expressions for the variation of the regulator functions,
\begin{equation}
\begin{split}
\delta_\omega R^{(1)\alpha \beta \gamma \delta} _k (x;\bar{g})
=&
-
\omega^\sigma \partial_\sigma
R^{(1)\alpha \beta \gamma \delta} _k (x;\bar{g})
+
R^{(1)\sigma \beta \gamma \delta} _k (x;\bar{g})
\partial_\sigma \omega^\alpha
+
R^{(1)\alpha \sigma \gamma \delta} _k (x;\bar{g})
\partial_\sigma \omega^\beta
\\
&+
R^{(1)\alpha \beta \sigma \delta} _k (x;\bar{g})
\partial_\sigma \omega^\gamma
+
R^{(1)\alpha \beta \gamma \sigma} _k (x;\bar{g})
\partial_\sigma \omega ^\delta
\n{eq:vR1}
\end{split}
\end{equation}
and
\beq
\delta_\omega R^{(2)} _{k \alpha \beta} (x;\bar{g})
=
-
\omega ^\sigma
\partial_\sigma R^{(2)} _{k \alpha \beta} (x;\bar{g})
-
R^{(2)} _{k \sigma \beta} (x;\bar{g})
\partial_\alpha \omega ^\sigma
-
R^{(2)} _{k \alpha \sigma} (x;\bar{g})
\partial_\beta \omega^\sigma
.
\n{eq:vR2}
\eeq

If the relations (\r{eq:vR1}) and (\r{eq:vR2}) are fulfilled, then the
action $S_k(\phi,\bar{g})$ is invariant under background field transformations.
Therefore, the regulator functions should have a tensor structure in order to
ensure the invariance.
Thus, taking into account the symmetry properties
presented in (\r{eq:symmetry}) we can propose the following
solutions for the regulator functions
\beq
R_k ^{(1) \alpha \beta \gamma \delta} (x;\bar{g})
=
\bar{g}^{\alpha \beta}(x)
\bar{g}^{\gamma \delta}(x)
R^{(1)} _k(\bar{\square} )
+
\big(\bar{g}^{\alpha \gamma}(x)
\bar{g}^{\beta \delta}(x)
+
\bar{g}^{\alpha \delta}(x)
\bar{g}^{\beta \gamma}(x)\big)
Q_k(\bar{\square})
\n{eq:nR1} \eeq and \beq \n{R2} R_{k \alpha \beta} ^{(2)}
(x;\bar{g}) = \bar{g}_{\alpha \beta} R^{(2)} _k(\bar{\square}) ,
\eeq where $R^{(1,2)} _k(\bar{\square})$ and $Q_k(\bar{\square} )$
are scalar functions and $\bar{\square}$ is the d'Alembertian
operator defined in terms of the covariant derivatives of the
background metric: \beq \bar{\square} = \bar{g}^{\mu \nu}
\bar{\nabla}_\mu \bar{\nabla}_\nu , \eeq with the metricity property
\beq \bar{\nabla}_\tau \bar{g}_{\mu\nu} = 0 . \eeq It is possible to
show that (\r{eq:nR1}) and \eq{R2} presents the same variational
structure of (\r{eq:vR1}) and \eq{eq:vR2}, respectively. By using
the inverse background metric variation \beq \delta_\omega
\bar{g}^{\mu \nu} (x) = - \omega^\sigma \partial_\sigma \bar{g}^{\mu
\nu}(x) + \bar{g}^{\mu \sigma}(x) \partial_\sigma \omega^\nu +
\bar{g}^{\sigma \nu}(x) \partial_\sigma \omega^\mu , \eeq we have
\beq
\delta_\omega R_k ^{(1) \alpha \beta \gamma \delta} (x;\bar{g})
&=&
\Big(
-\omega^\sigma
\partial_\sigma \bar{g}^{\alpha \beta}(x)
+
\bar{g}^{\alpha \sigma}(x)
\partial_\sigma \omega^\beta
+
\bar{g}^{\sigma \beta}(x)
\partial_\sigma \omega^\alpha
\Big) \bar{g}^{\gamma \delta}(x)R^{(1)} _k(\bar{\square})
\nn
\\
&+&
\bar{g}^{\alpha \beta}(x)
\Big(
-\omega^\sigma
\partial_\sigma \bar{g}^{\gamma \delta}(x)
+
\bar{g}^{\gamma \sigma}(x)
\partial_\sigma \omega^\delta
+
\bar{g}^{\sigma \delta}(x)
\partial_\sigma \omega^\gamma
\Big) R^{(1)} _k(\bar{\square})
\nn
\\
&+&
\Big(
-
\omega^\sigma
\partial_\sigma \bar{g}^{\alpha \gamma}(x)
+
\bar{g}^{\alpha \sigma}(x)
\partial_\sigma \omega^\gamma
+
\bar{g}^{\sigma \gamma}(x)
\partial_\sigma \omega^\alpha
\Big)
\bar{g}^{\beta \delta}(x)Q_k(\bar{\square})
\nn
\\
&+&
\bar{g}^{\alpha \gamma}(x)
\Big(
-
\omega^\sigma
\partial_\sigma \bar{g}^{\beta \delta}(x)
+
\bar{g}^{\beta \sigma}(x)
\partial_\sigma \omega^\delta
+
\bar{g}^{\sigma \delta}(x)
\partial_\sigma \omega^\beta
\Big)Q_k(\bar{\square})
\nn
\\
&+&
\Big(
-
\omega^\sigma
\partial_\sigma
\bar{g}^{\alpha \delta}(x)
+
\bar{g}^{\alpha \sigma}(x)
\partial_\sigma \omega^\delta
+
\bar{g}^{\sigma \delta}(x)
\partial_\sigma \omega^\alpha
\Big)
\bar{g}^{\beta \gamma}(x) Q_k (\bar{\square} )
\nn
\\
&+&
\bar{g}^{\alpha \delta}(x)
\Big(
-
\omega^\sigma
\partial_\sigma \bar{g}^{\beta \gamma}(x)
+
\bar{g}^{\beta \sigma}(x)
\partial_\sigma \omega^\gamma
+
\bar{g}^{\sigma \gamma}(x)
\delta_\sigma \omega^\beta
\Big) Q_k(\bar{\square})
\nn
\\
&-&
\bar{g}^{\alpha \beta}(x)
\bar{g}^{\gamma \delta}(x)
\omega^\sigma \partial_\sigma R^{(1)} _k(\bar{\square})
-
\bar{g}^{\alpha \gamma}(x)
\bar{g}^{\beta \delta}(x)
\omega^\sigma \partial_\sigma Q_k(\bar{\square})
\nn
\\
&-&
\bar{g}^{\alpha \delta}(x)
\bar{g}^{\beta \gamma}(x)
\omega^\sigma \partial_\sigma Q_k(\bar{\square}).
\eeq
The derivatives in metric tensor and in functions
$R^{(1)} _k(\bar{\square})$ and $Q_k(\bar{\square})$ can
be combined to obtain
\beq
\delta_\omega R_k ^{(1) \alpha \beta \gamma \delta} (x;\bar{g})
&=&
\Big(
\bar{g}^{\alpha \sigma}(x)
\partial_\sigma \omega^\beta
+
\bar{g}^{\sigma \beta}(x)
\partial_\sigma \omega^\alpha
\Big) \bar{g}^{\gamma \delta}(x)R^{(1)} _k(\bar{\square})
\nn
\\
&+&
\bar{g}^{\alpha \beta}(x)
\Big(
\bar{g}^{\gamma \sigma}(x)
\partial_\sigma \omega^\delta
+
\bar{g}^{\sigma \delta}(x)
\partial_\sigma \omega^\gamma
\Big)R^{(1)} _k(\bar{\square})
\nn
\\
&+&
\Big(
\bar{g}^{\alpha \sigma}(x)
\partial_\sigma \omega^\gamma
+
\bar{g}^{\sigma \gamma}(x)
\partial_\sigma \omega^\alpha
\Big)
\bar{g}^{\beta \delta}(x) Q_k(\bar{\square})
\nn
\\
&+&
\bar{g}^{\alpha \gamma}(x)
\Big(
\bar{g}^{\beta \sigma}(x)
\partial_\sigma \omega^\delta
+
\bar{g}^{\sigma \delta}(x)
\partial_\sigma \omega^\beta
\Big) Q_k(\bar{\square})
\nn
\\
&+&
\Big(
\bar{g}^{\alpha \sigma}(x)
\partial_\sigma \omega^\delta
+
\bar{g}^{\sigma \delta}(x)
\partial_\sigma \omega^\alpha
\Big)
\bar{g}^{\beta \gamma}(x) Q_k (\bar{\square})
\nn
\\
&+&
\bar{g}^{\alpha \delta}(x)
\Big(
+
\bar{g}^{\beta \sigma}(x)
\partial_\sigma \omega^\delta
+
\bar{g}^{\sigma \gamma}(x)
\delta_\sigma \omega^\beta
\Big) Q_k(\bar{\square})
\nn
\\
&-&
\omega^\sigma \partial_\sigma
\Big( \bar{g}^{\alpha \beta}(x)
\bar{g}^{\gamma \delta}(x) R^{(1)} _k(\bar{\square})
\Big)
-
\omega^\sigma \partial_\sigma
\Big( \bar{g}^{\alpha \gamma}(x)
\bar{g}^{\beta \delta}(x)Q_k(\bar{\square})
\Big)
\nn
\\
&-&
\omega^\sigma \partial_\sigma
\Big( \bar{g}^{\alpha \delta}(x)
\bar{g}^{\beta \gamma}(x) Q_k(\bar{\square})
\Big).
\n{eq:varF1}
\eeq
As a result, it is possible to see that (\r{eq:vR1}) is satisfied.

For the second function its variation can be expressed, after some algebra,
as
\beq
\delta_\omega
R_{k \alpha \beta} ^{(2)} (x;\bar{g})
&=&
-
 \omega^\sigma
\partial_\sigma \bar{g}_{\alpha \beta}
R^{(2)} _k(\bar{\square})
-
\partial_\alpha \omega^\sigma
\bar{g}_{\sigma \beta}(x) R^{(2)} _k(\bar{\square})
-
\partial_\beta \omega^\sigma
\bar{g}_{\alpha \sigma}(x)
R^{(2)} _k(\bar{\square})
\nn
\\
&-&
\omega^\sigma
\bar{g}_{\alpha \beta}(x)
\partial_\sigma R^{(2)} _k(\bar{\square}).
\eeq
The combination of derivatives in metric tensor
and in scalar function leads to
\beq
\delta_\omega
R_{k \alpha \beta} ^{(2)} (x;\bar{g})
=
-\omega ^\sigma
\partial_\sigma (\bar{g}_{\alpha\beta} (x) R^{(2)} _k(\bar{\square}))
-
\partial_\alpha \omega^\sigma
\bar{g}_{\sigma \beta}(x) R^{(2)} _k(\bar{\square})
-
\partial_\beta \omega^\sigma
\bar{g}_{\alpha \sigma}(x)
R^{(2)} _k(\bar{\square})
,
\eeq
which has the same structure as \eq{eq:vR2}.

Finally, the scale-dependent regulator Lagrangian density (\r{eq:lagr})
in terms of \eq{eq:nR1} 
and \eq{R2} reads
\begin{equation}
\begin{split}
\mathcal{L}_k(\phi ,\bar{g})&=
\frac{1}{2}h_{\alpha\beta}(x)
\Big[\bar{g}^{\alpha \beta}(x)
\bar{g}^{\gamma \delta}(x)
R^{(1)} _k(\bar{\square})
\\
& +
\big(
\bar{g}^{\alpha \gamma}(x)
\bar{g}^{\beta \delta}(x)
+
\bar{g}^{\alpha \delta}(x)
\bar{g}^{\beta \gamma}(x)
\big)
Q_k(\bar{\square})\Big] h_{\gamma\delta}(x)
\\
&+
\bar{C}^\alpha(x)\bar{g}_{\alpha \beta}(x)
R^{(2)} _k (\bar{\square})C^{\beta}(x),
\end{split}
\end{equation}
which maintains the background field symmetry, $\de_\om S_k(\phi,\bar{g}) = 0$.

\section{Gauge dependence of effective average  action}
\label{sect4}

In order to understand the gauge invariance and gauge dependence
problems in the background field method, we shall consider the
generating functionals  of Green functions \beq Z_{k\Psi}(J,\bar{g})
&=& \int d\phi \exp\Big\{ \frac{i}{\hbar}[S_0(h+\bar{g})+
\hat{R}(\phi ,\bar{g})\Psi(\phi,\bar{g})
+S_{k}(\phi,\bar{g})+J\phi]\Big\} \nn
\\
&=& \int d\phi \exp\Big\{ \frac{i}{\hbar}[S_{kFP}(\phi
,\bar{g})+J\phi] \Big\} = \exp\Big\{
\frac{i}{\hbar}W_{k\Psi}(J,\bar{g}) \Big\}, \n{eq:Z} \eeq where \beq
\Psi(\phi ,\bar{g}) = \int
dx\sqrt{-\bar{g}(x)}\bar{C}^\alpha(x)\chi_\alpha(x;h,\bar{g}) \eeq
is the fermionic gauge fixing functional and \beq
\hat{R}(\phi,\bar{g}) = \int dx \frac{\delta_r}{\delta
\phi^i(x)}R^i(x;\phi,\bar{g}) \eeq is the generator of BRST
transformations (\r{eq:BRSTcon}).

As far as we saw in the previous section, the regulator action
(\r{eq:act}) does not depend on the gauge $\Psi(\phi ,\bar{g})$.
Now, we  shall consider another choice of gauge fixing functional
$\Psi \to \Psi + \delta\Psi$ and set $J=0$ in (\r{eq:Z}). Thus,
\beq
Z_{k\Psi+\delta\Psi}(\bar{g}) = \int d\phi \exp \Big\{
\frac{i}{\hbar}[S_{k FP}(\phi,\bar{g}) + \hat{R}(\phi
,\bar{g})\delta\Psi(\phi ,\bar{g})] \Big\} =
\exp\Big\{\frac{i}{\hbar}W_{k\Psi+\delta\Psi}(\bar{g})\Big\},
\n{eq:Z1}
\eeq
where
\beq \delta\Psi = \delta\Psi(\phi ,\bar{g}) =
\int dx \sqrt{-\bar{g}(x)} \bar{C}^\alpha(x)\delta\chi_\alpha
(h,\bar{g}). \n{eq:Zk} \eeq

We will try to compensate the additional term $\hat{R}\delta\Psi$ in
(\r{eq:Z1}). To do this, we change the variables in the functional
integral related to the symmetries of action $S_{FP}(\phi
,\bar{g})$,  namely the BRST symmetry and the background gauge
invariance. First, we shall consider the BRST symmetry
(\r{eq:BRST}), but trading the constant parameter $\lambda$ by a
functional $\Lambda = \Lambda(\phi , \bar{g})$. The variation of
(\r{eq:act}) under such transformation is the following \beq
\delta_B S_k (\phi,\bar{g}) = \int d^4x\sqrt{-\bar{g}(x)} \Big\{
\delta_B \mathcal{L}_k(\phi ,\bar{g}) \Big\}, \eeq where
\begin{equation}
\begin{split}
\delta_B \mathcal{L}_k
=& \,\,
\frac{1}{2}\delta_B h_{\alpha\beta}(x)
R^{(1)\alpha\beta\gamma\delta} _k (x;\bar{g})
h_{\gamma\delta}(x)
+
\frac{1}{2}h_{\alpha\beta}(x)
R^{(1)\alpha\beta\gamma\delta} _k (x;\bar{g})
\delta_B h_{\gamma\delta}(x)
\nn
\\
&+
\delta_B \bar{C}^\alpha(x)
R^{(2)} _{k\alpha\beta}(x;\bar{g})
C^\beta(x)
+
\bar{C}^\alpha(x)
R^{(2)} _{k\alpha\beta}(x;\bar{g})
\delta_B C^\beta(x)
\end{split}
\end{equation}
After some algebra, $\delta_B\mathcal{L}_k(\phi,\bar{g})$ reads as
\begin{equation}
\begin{split}
\delta_B \mathcal{L}_k
=& \,\,
-\frac{1}{2}
\big(C^\sigma(x)\partial_\sigma g_{\alpha\beta}(x)
+
g_{\alpha\sigma}(x)\partial_\beta C^\sigma(x)
+
g_{\sigma\beta}(x)\partial_\alpha C^\sigma(x)\big)\Lambda
R^{(1)\alpha\beta\gamma\delta} _k (x;\bar{g})
h_{\gamma\delta}(x)
\nn
\\
&-
\frac{1}{2}h_{\alpha\beta}(x)
R^{(1)\alpha\beta\gamma\delta} _k (x;\bar{g})
\big(C^\sigma(x)\partial_\sigma g_{\gamma\delta}(x)
+
g_{\gamma\sigma}(x)\partial_\delta C^\sigma(x)
+
g_{\sigma\delta}(x)\partial_\gamma C^\sigma(x)\big)\Lambda
\nn
\\
&+
B^\alpha(x)\Lambda
R^{(2)} _{k\alpha\beta}(x;\bar{g})
C^\beta(x)
+
\bar{C}^\alpha(x)
R^{(2)} _{k\alpha\beta}(x;\bar{g})
C^\sigma(x) \partial_\sigma C^\beta(x)\Lambda .
\end{split}
\end{equation}

From the above expression, it is clear that the action $S_k(\phi
,\bar{g})$ is not invariant under BRST transformations $\de_B
S_k(\phi ,\bar{g}) \neq 0$.  The Jacobian $J_1$ of such
transformation can be obtained in the standard way \beq J_1 = \exp
\Big\{ \int dx \Big[ \frac{\delta (\delta_B
h_{\alpha\beta}(x))}{\delta h_{\alpha\beta}(x)} - \frac{\delta
(\delta_B C^\alpha(x))}{\delta C^\alpha(x)} - \frac{\delta (\delta_B
\bar{C}^\alpha(x))}{\delta \bar{C}^\alpha(x)} \Big] \Big\}, \eeq
where the functional derivatives are
\begin{align}
\frac{\delta (\delta_B h_{\alpha\beta}(x))}{\delta h_{\alpha\beta}(x)}
=&
-\frac{D(D+1)}{2}\delta(0)C^\sigma(x)\partial_\sigma\Lambda(\phi,\bar{g})
-
\frac{(D+1)(D-2)}{2}\delta(0)\partial_\sigma C^\sigma(x)\Lambda(\phi,\bar{g})
\nn
\\
&-\Big[
C(x)^\sigma\partial_\sigma g_{\alpha\beta}(x)
+
g_{\alpha\sigma}(x)\partial_\beta C^\sigma(x)
+
g_{\sigma\beta}(x)\partial_\alpha C^\sigma(x)
\Big]
\frac{\delta\Lambda(\phi,\bar{g})}{\delta h_{\alpha\beta}(x)},
\\
\frac{\delta (\delta_B C^\alpha(x))}{\delta C^\alpha(x)}
=& \,\,
(D+1)\delta(0)\partial_\sigma C^\sigma(x)\Lambda(\phi,\bar{g})
+
D\delta(0) C^\sigma(x)\partial_\sigma\Lambda(\phi,\bar{g})
\nn
\\
&+
C^\sigma(x)\partial_\sigma C^\alpha(x)
\frac{\delta\Lambda(\phi,\bar{g})}{\delta C^\alpha(x)},
\\
\frac{\delta (\delta_B \bar{C}^\alpha(x))}{\delta \bar{C}^\alpha(x)}
=& \,\,
B^\alpha(x)\frac{\delta\Lambda(\phi,\bar{g})}{\delta\bar{C}^\alpha(x)} .
\end{align}

It is possible to choose a regularization scheme such  that
$\delta(0)=0$.  As a result, the Jacobian for BRST transformations
is
\begin{equation}
J_1 =
\exp \Big\{
\int dx\Big[
R_{\alpha\beta} ^{(h)}(x;\phi,\bar{g})
\frac{\delta\Lambda(\phi,\bar{g})}{\delta h_{\alpha\beta}(x)}
-
R_{(C)} ^\alpha(x;\phi,\bar{g})
\frac{\delta\Lambda(\phi,\bar{g})}{\delta C^\alpha(x)}
-
R_{(\bar{C})} ^\alpha(x;\phi,\bar{g})
\frac{\delta\Lambda(\phi,\bar{g})}{\delta\bar{C}^\alpha(x)}
\Big]
\Big\}.
\end{equation}
where (\r{eq:BRSTcon}) is used.

It is also interesting to consider the background gauge
transformation related to expressions (\r{eq:backgauge}). As far as
the regulator functions transform as (\r{eq:vR1}) and (\r{eq:vR2}),
the action $S_k(\phi ,\bar{g})$  is invariant under such
transformation.  But now, instead of functions
$\omega=\omega^\sigma(x)$ we shall consider the functional
$\Omega^\sigma = \Omega^\sigma (x, \phi, \bar{g})$. The action
(\r{eq:act}) remains invariant, and the corresponding Jacobian of
this transformation can be obtained as before \beq J_2 = \exp \Big\{
\int dx \frac{\delta (\delta_\Omega h_{\alpha\beta}(x))}{\delta
h_{\alpha\beta}(x)} - \frac{\delta (\delta_\Omega
C^\alpha(x))}{\delta C^\alpha(x)} - \frac{\delta (\delta_\Omega
\bar{C}^\alpha(x))}{\delta \bar{C}^\alpha(x)} \Big\}, \eeq with the
following functional derivatives:
\begin{align}
\frac{\delta (\delta_\Omega h_{\alpha\beta}(x))}{\delta h_{\alpha\beta}(x)}
=& \,\,
\frac{(D+1)(D-2)}{2}\delta(0)\partial_\sigma\Omega^\sigma(x,\phi,\bar{g})
\nn
\\
&-
\big(
\partial_\sigma h_{\alpha\beta}(x)+h_{\alpha\sigma}(x)\partial_\beta +
h_{\sigma\beta}(x)\partial_\alpha
\big)
\frac{\delta\Omega^\sigma(x,\phi,\bar{g})}{\delta h_{\alpha\beta}(x)},
\\
\frac{\delta (\delta_\Omega \bar{C}^\alpha(x))}{\delta
\bar{C}^\alpha(x)} =& \,\,
(D+1)\delta(0)\partial_\sigma\Omega^\sigma(x,\phi,\bar{g}) -
\frac{\delta\Omega^\sigma(x,\phi,\bar{g})}{\delta\bar{C}^\alpha(x)}
\partial_\sigma\bar{C}^\alpha(x) \nn
\\
&-
\bar{C}^\sigma(x)\partial_\sigma\frac{\delta\Omega^\alpha(x,\phi,\bar{g})}
{\delta\bar{C}^\alpha(x)},
\\
\frac{\delta(\delta_\Omega C^\alpha(x))}{\delta C^\alpha(x)} =&\,\,
(D+1)\delta(0)\partial_\sigma\Omega^\sigma(x,\phi,\bar{g}) -
\frac{\delta\Omega^\sigma(x,\phi,\bar{g})}{\delta C^\alpha(x)}
\partial_\sigma C^\alpha(x) \nn
\\
&- C^\sigma(x)\partial_\sigma\frac{\delta\Omega^\alpha(x,\phi,
\bar{g})}{\delta C^\alpha(x)}.
\end{align}

As before, $\delta(0)=0$. Thus, the Jacobian for background gauge
transformations reads
\begin{equation}
\begin{split}
J_2 =& \,\,
\exp \Big\{
\int dx \Big[
-\big(
\partial_\sigma h_{\alpha\beta}(x)+h_{\alpha\sigma}(x)\partial_\beta +
h_{\sigma\beta}(x)\partial_\alpha\big)
\frac{\delta\Omega^\sigma(x,\phi,\bar{g})}{\delta h_{\alpha\beta}(x)}
\\
&+
\frac{\delta\Omega^\sigma(x,\phi,\bar{g})}{\delta\bar{C}^\alpha(x)}
\partial_\sigma\bar{C}^\alpha(x) +
\bar{C}^\sigma(x)\partial_\sigma\frac{\delta\Omega^\alpha(x,\phi,\bar{g})}
{\delta\bar{C}^\alpha(x)} +
\frac{\delta\Omega^\sigma(x,\phi,\bar{g})}{\delta
C^\alpha(x)}\partial_\sigma C^\alpha(x)
\\
&+ C^\sigma(x)\partial_\sigma\frac{\delta\Omega^\alpha(x,
\phi,\bar{g})}{\delta C^\alpha(x)} \Big] \Big\}.
\end{split}
\end{equation}

If it is possible to fulfill the condition \beq
J_1J_2\exp\Big\{\frac{i}{\hbar}\int dx
[\hat{R}(\phi,\bar{g})\delta\Psi(\phi,\bar{g}) + \delta_B
S_k(\phi,\bar{g})] \Big\}=1 \n{eq:condition} \eeq the generating
vacuum functional $Z_{k\Psi}(\bar{g})$ does not depend on the gauge
fixing functional $\Psi$. In order to verify that, let us expand
the functionals $\Lambda$ and $ \Omega$, with Grassmann parity
$\varepsilon(\Lambda)=1$ and $\varepsilon(\Omega)=0$ and ghost
numbers $\mathrm{gh}(\Lambda)=-1$ and $\mathrm{gh}(\Omega)=0$ in the
lower power of ghost fields \beq \Lambda = \Lambda^{(1)} +
\Lambda^{(3)}, \qquad \Omega ^\sigma = \Omega ^{\sigma(0)} + \Omega
^{\sigma(2)}, \eeq where
\begin{align}
\Lambda^{(1)} &= \int dx
\bar{C}^\alpha(x)\lambda^{(1)} _\alpha (x,h,\bar{g}),
\\
\Lambda^{(3)} &= \int dx
\frac{1}{2}\bar{C}^\alpha(x)\bar{C}^\beta(x)
\lambda^{(3)} _{\alpha\beta\gamma}(x,h,\bar{g})C^\gamma(x),
\\
\Omega^{\sigma(0)} (x) &=
\Omega^{\sigma(0)}(x,h,\bar{g}),
\\
\Omega^{\sigma(2)} (x,h,\bar{g}) &=
\bar{C}^{\alpha}(x)
\omega^{\sigma(2)} _{\alpha\beta}(x,h,\bar{g})C^\beta(x).
\end{align}

The terms that vanish in (\r{eq:condition}) and do not depend on
ghost fields,  leads to \beq \n{omega0} \big(
\partial_\sigma h_{\alpha\beta}(x)
+ h_{\alpha\sigma}(x)\partial_\beta +
h_{\sigma\beta}(x)\partial_\alpha \big)
\frac{\delta\Omega^{\sigma(0)}(x,h,\bar{g})}{\delta
h_{\alpha\beta}(x)} =0 . \eeq Analyzing the terms which are linear
in the anti-ghost fields,  which contains the auxiliary fields
$B(x)$, we obtain \beq \lambda^{(1)} _\alpha (x,h,\bar{g}) =
\frac{i}{\hbar}\delta\chi_\alpha(x,h,\bar{g}) \eeq and \beq \n{L3}
\lambda^{(3)} _{\alpha\beta\gamma}(x,h,\bar{g}) = \lambda^{(1)}
_\alpha(h,\bar{g})R^{(2)} _{k\beta\gamma}(x;\bar{g}) - \lambda^{(1)}
_\beta(h,\bar{g})R^{(2)} _{k\alpha\gamma}(x;\bar{g}), \eeq where
\beq \lambda^{(1)} _\alpha(h,\bar{g}) = \int dx \lambda^{(1)}
_\alpha(x,h,\bar{g}). \eeq

Now, the vanishing terms with structure $\bar{C}(x)C(x)$  can be related to the
second order of $\Omega^\sigma(x,h,\bar{g})$ functional leading to a
differential equation for $\omega^{\sigma(2)} _{\alpha\beta}(x,h,\bar{g})$,
\beq
\partial_\sigma\omega^{\sigma(2)} _{\alpha\beta}(x,h,\bar{g})
&=&
\frac{i}{2\hbar}\Big[
\partial_\beta g_{\tau\sigma}(x)\lambda_\alpha ^{(1)}(h,\bar{g})
R^{(1)\tau\sigma\gamma\delta}(x;\bar{g})h_{\gamma\delta}(x)
\nn
\\
&+& h_{\tau\sigma}(x)R^{(1)\tau\sigma\gamma\delta}
(x;\bar{g})\partial_\beta g_{\gamma\delta}(x) \lambda^{(1)}
_\alpha(h,\bar{g}) \Big] \,. \eeq

From Eq.~\eqref{omega0} since $ \Omega^{\sigma(0)}(x,h,\bar{g}) $ is
an arbitrary function we can not have just one particular solution.
In addition, the $ \la^{(3)}_{\alpha\beta\gamma} $ relation in
\eqref{L3} creates non-local term of structure $B \bar{C} \bar{C}  C
C$ which can be only  eliminated if we consider new functional
$\Lambda $ of higher orders in ghost fields. Even so, this process
would repeat endlessly. The only case left for us is to consider the
simple solution when $ \Om^\si = 0 $ and $ \Lambda = \Lambda^{(1)} $
we have the result
\beq
Z_{k\Psi + \de \Psi}(\bar{g}) &=&
\int d\phi
\exp\Big\{
\frac{i}{\hbar}
[S_{kFP}(\phi ,\bar{g})
+ \de S_{kFP}(\phi )]
\Big\} \, ,
\nn \\
Z_{k\Psi }(\bar{g}) &\neq& Z_{k\Psi + \de \Psi}(\bar{g})
\, .
\eeq
As final result, the vacuum functional in the FRG approach for
gravity theories depends on the gauge fixing even on-shell,  which
leads to a gauge dependent $S$-matrix.

\section{Conclusions and Perspectives}
\label{sect5}

We explored the problem of gauge invariance and the gauge-fixing
dependence using the background field formalism, for gravity theories
in the FRG framework. It was shown that the background field
invariance is achieved when the regulator functions are choosen to
have the tensor structure. Nevertheless, even in this case the on-shell
gauge dependence cannot be cured in the standard FRG approach which
we dealt with. In this respect the situation is qualitatively the same as in
the Yang-Mills theories, as it was discussed in \cite{Lavrov:2012xz}.

The on-shell gauge dependence takes place due to the fact that the
regulator action (\r{eq:act}) is not BRST-invariant. It turns out
that this is a fundamentally important feature, that can be changed
only by trading the standard and conventional FRG framework to an
alternative one, which is based on the use of composite operators
for constructing the regulator action. Unfortunately, until now
there is no way to perform practical calculations in this
alternative formulation. For this reason, taking our present results
into account, it remains unclear whether the quantum gravity results
obtained within the FRG formalism can have a reasonable physical
interpretation. One can expect that all predictions of this
formalism will depend on an arbitrary choice of the gauge fixing.
Thus, one can, in principle, provide any desirable result, but the
value of this output is not clear. Alternatively, there should be
found some physical reason to claim that one special gauge fixing is
``correct'' or ``preferred'' for some reason, but at the moment it
is unclear how this reason can look like, since the original theory
is gauge (diffeomorphism) invariant.


It is worthwhile to comment on the effect of the on-shell gauge
fixing dependence on the non-universality of the renormalization
group flow in the framework of the Functional Renormalization Group
approach. First of all, the complete understanding of this issue
requires the detailed analysis and, in principle, explicit
calculations in the sufficiently general model of quantum gravity.
This calculation is a separate problem that is beyond the scope
of the present work. At the same time, we can give some strong
arguments in favor of such a gauge fixing dependence for both
renormalization group flow and the fixed point, which is obtained
from this flow.

One can classify the models of quantum gravity into two classes.
The first class is the quantum general relativity, where the
renormalization group equation for the Newton constant $G$ does not
exist at the perturbative level, at least without the cosmological
constant term. In the presence of the cosmological constant there
is a perturbative renormalization group equation for the Newton
constant, but it is strongly gauge-fixing dependent, as discussed
e.g. in \cite{frts82} and more recently, with a very general
calculations, in \cite{JDG-QG} (see more references therein).
Let us add that in the Functional Renormalization Group approach
the renormalization group flow for $G$ is extracted from the
quadratic divergences, and for this reason the corresponding results
do not have the perturbative limit. On the other hand, the quadratic
divergences are also gauge-fixing dependent  \cite{JDG-QG} and,
furthermore, suffer from the total ambiguity in defining the cut-off
parameter $\Om$. For instance, if we start from the total expression
for the one-loop divergences in the Schwinger-DeWitt formalism,
\beq
\Ga^{(1)}_{div} \,=\,-\frac12
\int d^4x \sqrt{g}\bigg\{ \frac12\,A_0 \Om^4 + A_1 \Om^2
+ A_2 \log \Big(\frac{\Om^2}{\mu^2}\Big)\bigg\},
\label{A012}
\eeq
where $A_0$, $A_1$ and $A_2$ are the algebraic sums of the
contributions of quantum metric and ghosts, given by the
functional traces of the coincidence limits of the corresponding
 Schwinger-DeWitt coefficients. Without the cosmological
constant only $A_1$ has the term linear in the scalar curvature
$R$, and hence only this term can contribute to the renormalization
group flow for $G$. Now, changing the cut-off as
\beq
\Om^2
\quad  \longrightarrow \quad
\Om^2 + \al R,
\label{Omega}
\eeq
with an arbitrary coefficient $\al$,
we observe that the quartic and logarithmic divergences do not
change, while the quadratic ones modify according to\footnote{One
of the authors (I.Sh.) is grateful to M. Asorey for explaining this
point.}
\beq
A_1
\quad  \longrightarrow \quad
A_1 \,+\, \al A_0 R,
\label{A1}
\eeq
making it completely ambiguous. Thus, in our opinion, it is difficult
to take the results extracted from $A_1$ as a physical prediction,
even without gauge fixing ambiguity.

The theories of quantum gravity of the second kind are those with
higher derivatives, being fourth derivative \cite{Stelle77} models,
 the superrenormalizable models with six or more derivatives
 \cite{highderi} or the nonlocal superrenormalizable models. In
 all these cases the renormalization group flow in the Functional
 Renormalization Group has a well-defined perturbative limit (see
 e.g. \cite{Percacci-2007}) and hence it should be supposed that
 the invariance or non-invariance of the renormalization group flow
 should be the same in both usual perturbative and Functional
 Renormalization Group approaches. On the other hand, without
 the on-shell invariance condition there is absolutely no way to
 consider the renormalization group flows in any one of these
 models in a consistent way. This issue was analyzed in detail
 in \cite{frts82,avram-tes,a,highderi} and \cite{QG-betas}, thus
 there is no point to repeat these considerations here. Thus, we can
 conclude that in those versions of Functional Renormalization
 Group where one can expect a minimally reasonable interpretation
 of the application to quantum gravity, it is difficult to expect
 the gauge-invariant flow in the framework of the Functional
 Renormalization Group, if this approach is not reformulated in
 a new invariant way.

Finally, we mention two recent papers. In \cite{LMS} it was shown
that the renormalization group flow in quantum general relativity
should be compatible with the Ward identities related to the
possibility to change the conformal frame. Certainly, this is not
the symmetry that we discussed in the previous sections. Also,
even if the mentioned compatibility is achieved, this does not
make the whole approach sufficiently unambiguous, as we
explained above. However, qualitatively, the importance of
the Ward identities to the consistency of the renormalization
group flow is certainly coherent with our analysis. On the other
hand, Ref.~\cite{Alwis} there was proposed a new formulation
of the Functional Renormalization Group for quantum gravity.
In this new version, there is not one scale parameters, but two:
one for regularizing the UV sector and another one for regularizing
the IR sector. The idea is certainly interesting for quantum gravity
since the two limits in all known models are very much different.
However, since the UV part is essentially the same as in the
standard formulation, it is difficult to expect that the gauge
fixing dependence on-shell in this new formulation will be a
smaller problem than it is in the standard versions with one
scale parameter. We can state that the preliminary analysis of
this model confirms this qualitative conclusion. The full
consideration is possible, but would be quite involved and
is beyond the scope of the present work.


The results of the considerations which we described above make
more interesting the discussion of the possible ways to solve the
problem of on-shell gauge fixing dependence, such as the ones
suggested in \cite{Morris,Arnone,Rosten} or in \cite{Lavrov:2012xz}.
In our opinion, the last approach is more transparent and physically
motivated, but (as we have already mentioned above) there is no
well-developed technique of using it for practical calculations.

\section*{Acknowledgments}

E.A.R. and V.F.B. are grateful to  Coordena\c{c}\~{a}o de
Aperfei\c{c}oamento de Pessoal de N\'{i}vel Superior - CAPES for
supporting their Ph.D. and Ms. projects. P.M.L. is grateful to the
Department of Physics of the Federal University of Juiz de Fora (MG,
Brazil) for warm hospitality during his long-term visit, when this
work has been initiated.  The work of P.M.L. was partially supported
by  the Ministry of Science and Higher Education of the Russian
Federation, grant  3.1386.2017 and by the RFBR grant 18-02-00153.
I.L.Sh. was partially supported by Conselho Nacional de
Desenvolvimento Cient\'{i}fico e Tecnol\'{o}gico - CNPq under the
grant 303635/2018-5 and Funda\c{c}\~{a}o de Amparo \`a Pesquisa de
Minas Gerais - FAPEMIG under the project APQ-01205-16.



\begin{thebibliography}{99}
%


\bibitem{Weinberg-79}
S. Weinberg, {\it Ultraviolet divergences in quantum theories of
gravitation}.
(In S. W. Hawking; W. Israel (eds.). General Relativity:
An Einstein centenary survey. Cambridge University Press, 1979).

\bibitem{NiederReuter}
 M.~Niedermaier and M.~Reuter,
{\it The Asymptotic Safety Scenario in Quantum Gravity,}
Living Rev. Rel.  {\bf 9} (2006) 5, gr-qc/0610018.

\bibitem{Percacci-2007}  R.~Percacci,
{\it Asymptotic Safety,}
(In D. Oriti - Editor: Approaches to quantum gravity, 111-128,
Cambridge University Press, 2007),
arXiv:0709.3851.

\bibitem{Stelle77} K.S.~Stelle,
{\it Renormalization of Higher Derivative Quantum Gravity},
 Phys. Rev. {\bf D16} (1977)  953.

\bibitem{HDQG} F.~de~O. Salles, and I.L.~Shapiro,
{\it Do we have unitary and (super)renormalizable quantum gravity
below the Planck scale?}.
Phys. Rev. {\bf D89} (2014) 084054;
Erratum: Phys. Rev. {\bf D90} (2014) 129903,
hep-th/1401.4583;
\\
A.M.~Pelinson, F.~de~Oliveira Salles, and I.L.~Shapiro, {\it
Gravitational waves and perspectives for quantum gravity}, Mod.\
Phys.\ Lett. {\bf A29} (2014) 1430034 (Brief Review),
gr-qc/1410.2581;
\\
P.~Peter, F.~de~Oliveira Salles, and I.L.~Shapiro,
{\it On the ghost-induced instability on de Sitter background},
Phys. Rev. {\bf D97} (2018)  064044,
gr-qc/1801.00063;
\\
A.~Salvio, {\it Metastability in Quadratic Gravity,} Phys. Rev. {\bf
D99} (2019) 103507,
arXiv:1902.09557;
\\
S.~Castardelli dos Reis, G.~Chapiro and I.~L.~Shapiro,
{\it Beyond the linear analysis of stability in higher derivative gravity
with the Bianchi-I metric,}
Phys. Rev. {\bf D100} 066004 (2019)
arXiv:1903.01044.

\bibitem{salstr}  A.~Salam and J.~Strathdee,
{\it Remarks on High-energy Stability and Renormalizability of
Gravity Theory}, Phys. Rev.  {\bf D18} (1978)  4480.

\bibitem{frts82} E.S.~Fradkin and  A.A.~Tseytlin,
{\it Renormalizable asymptotically free quantum theory of gravity},
Nucl. Phys. {\bf B201} (1982) 469.

\bibitem{avbar86} I.G.~Avramidi and A.O.~Barvinsky,
{\it Asymptotic Freedom In Higher Derivative Quantum Gravity}
Phys. Lett. {\bf B159} (1985) 269.

\bibitem{Gauss} G.~de~Berredo-Peixoto and I.L.~Shapiro,
{\it Higher derivative quantum gravity with Gauss-Bonnet term},
Phys. Rev. {\bf D71} (2005) 064005, hep-th/0412249.

\bibitem{Tomboulis-77} E.~Tomboulis,
{\it 1/N Expansion and Renormalization in Quantum Gravity},
Phys. Lett. {\bf B70} (1977)  361;
{\it Renormalizability and Asymptotic Freedom in Quantum Gravity},
Phys. Lett. {\bf B97} (1980)  77;
{\it Unitarity in Higher Derivative Quantum Gravity},
Phys. Rev. Lett. {\bf 52} (1984) 1173.

\bibitem{antomb} I.~Antoniadis and E.T.~Tomboulis,
{\it Gauge Invariance And Unitarity In Higher Derivative Quantum
Gravity}, Phys. Rev. {\bf D33} (1986)  2756.

\bibitem{Johnston} D.A.~Johnston,
{\it Sedentary Ghost Poles In Higher Derivative Gravity},
Nucl. Phys. {\bf B297} (1988) 721.

\bibitem{CodPer-06} A.~Codello and R.~Percacci,
{\it Fixed points of higher derivative gravity},
Phys. Rev. Lett.  {\bf 97} (2006) 221301, hep-th/0607128;
A.~Codello, R.~Percacci, L.~Rachwal and A.~Tonero,
{\it Computing the Effective Action with the Functional Renormalization
Group}, Eur. Phys. J. {\bf C76} (2016) 226, hep-th/1505.03119.

\bibitem{CodPerRah-2009}
  A.~Codello, R.~Percacci and C.~Rahmede,
{\it Investigating the Ultraviolet Properties of Gravity with a
Wilsonian Renormalization Group Equation,}
Annals Phys.  {\bf 324} (2009) 414,
arXiv:0805.2909.

\bibitem{Lavrov:2012xz} P.M.~Lavrov and I.L.~Shapiro,
{\it On the Functional Renormalization Group approach for
Yang-Mills fields,}
JHEP {\bf 1306} (2013) 086,
arXiv:1212.2577.

\bibitem{Lav(in)}
P.M. Lavrov, {\it Gauge (in)dependence and background field
formalism,} Phys.\ Lett. {\bf B791} (2019) 29; arXiv:1805.02149.

\bibitem{FRG-ELT}
P.M. Lavrov, E. A. dos Reis, T. de Paula Netto and  I.L.Shapiro,
{\it Gauge invariance of the background average effective action,}
Eur. Phys. J. {\bf C79} (2019)  661,
arXiv:1905.08296.

\bibitem{highderi} M.~Asorey, J.L.~L\'opez and I.L.~Shapiro,
{\it Some remarks on high derivative quantum gravity}, Int. J. Mod.
Phys. {\bf A12} (1997)  5711, hep-th/9610006.

\bibitem{QG-betas} L.~Modesto, L.~Rachwa\l \
and I.L.~Shapiro,
{\it Renormalization group in superrenormalizable quantum gravity,}
Eur. Phys. J. {\bf C78} (2018) 555,
arXiv:1704.03988.

\bibitem{YMtheories}
C. N. Yang and R. L. Mills, {\it Conservation of Isotopic Spin and
Isotopic Gauge Invariance,} Phys. Rev. {\bf 96} (1954) 191.

\bibitem{Jackiw}
R. Jackiw, {\it Functional evaluation of the effective potential,}
Phys. Rev. {\bf D9} (1974) 1686.

\bibitem{Dolan}
L. Dolan and R. Jackiw, {\it Gauge invariant signal for gauge
symmetry breaking,} Phys. Rev. {\bf D9} (1974) 2904.

\bibitem{Nielsen}
N.K. Nielsen, {\it On the gauge dependence of spontaneous symmetry
breaking in gauge theories,} Nuc. Phys. {\bf B101} (1975) 173.

\bibitem{Fukuda}
R. Fukuda and T. Kugo, {\it Gauge invariance in the effective action
and potential,} Phys. Rev. {\bf D13} (1976) 3469.

\bibitem{LavTyu}
P.M. Lavrov and I.V. Tyutin, {\it On the generating functional for
the vertex functions in Yang-Mills theories,} Sov. J. Nucl. Phys.
{\bf 36} (1981) 474.

\bibitem{Voronov}
B.L. Voronov, P.M. Lavrov and I.V. Tyutin, {\it Canonical
transformations and gauge dependence in general gauge theories,}
Sov. J. Nucl. Phys. {\bf 36} (1982) 292.

\bibitem{Wetterich}
C. Wetterich, {\it Average action and the renormalization group
equations,} Nuc. Phys. {\bf B352} (1991) 529.

\bibitem{Wetterich2}
C. Wetterich, {\it Exact evolution equation for the effective
potential,} Phys. Lett. {\bf B301} (1993) 90, arXiv:1710.05815.

\bibitem{Berges}
J. Berges, N. Tetradis and C. Wetterich, {\it Non-perturbative
renormalization flow in quantum field theory and statistical
physics,} Phys. Rept. {\bf 363} (2002) 223, hep-ph/0005122.

\bibitem{Bagnuls}
C. Bagnuls and C. Berviller, {\it Exact renormalization group
equations: an introductory review,}  Phys. Rept. {\bf 348}
(2001) 91, hep-th/0002034.

\bibitem{Polonyi}
J. Polonyi, {\it Lectures on the functional renormalization group method,}
 Central Eur. J. Phys. {\bf 1} (2003) 1, hep-th/0110026.

\bibitem{Pawlowski}
J.M. Pawlowski, {\it Aspects of the functional renormalization
group,} Annals Phys. {\bf 322} (2007) 2831, hep-th/0512261.

\bibitem{Delamotte}
B. Delamotte, {\it An introduction to the nonperturbative renormalization group,}
 Lect. Notes Phys. {\bf 852} (2012) 49, cond-mat/0702365.

\bibitem{Rosten}
O.J. Rosten, {\it Fundamentals of the exact renormalization group,}
Phys. Rep. {\bf 511} (2012) 177, arXiv:1003.1366.

\bibitem{Morris} T.R. Morris,
{\it A gauge invariant exact renormalization group. 2.} JHEP {\bf
12} (2000) 012, hep-th/0006064.

\bibitem{Arnone} S. Arnone, T.R. Morris, and O.J. Rosten,
{\it A generalized manifestly gauge invariant exact renormalization
group for SU(N) Yang-Mills,} Eur. Phys. J.  {\bf C50} (2007) 467,
hep-th/0606181.

\bibitem{Branchina}
V. Branchina, K.A. Meissner and G. Veneziano,
{\it The prize of an exact, gauge invariant RG flow equation,}
 Phys. Lett. {\bf B 574} (2003) 319, hep-th/0309234.

\bibitem{Pawlowski2}
J.M. Pawlowski, {\it Geometrical effective action and Wilsonian flows},
hep-th/0310018.

\bibitem{Vilkovisky}
G.A. Vilkovisky, in {\it B.S. DeWitt Sixtieth Anniversary Volume, S.}
Christensen eds., Hilger, Bristol U.K. (1983).

\bibitem{Vilkovisky2}
G.A. Vilkovisky, in {\it The unique effective action in quantum field theory,}
 Nucl. Phys. {\bf B234} (1984) 124.


\bibitem{DeWitt-BFM}
B.S. DeWitt, {\it Quantum theory of gravity. II.
The manifestly covariant theory,}  Phys. Rev. {\bf 162} (1967) 1195.

\bibitem{Arefeva}
I.Ya. Arefeva, L.D. Faddeev and A.A. Slavnov,
{\it Generating functional for the s matrix in gauge theories,}
Theor. Math. Phys. {\bf 21} (1975) 1165 ( Teor. Mat. Fiz.
{\bf 21}(1974) 311-321).

\bibitem{Abbott}
L.F. Abbott, {\it The background field method beyond one loop,}
Nucl. Phys. {\bf B185} (1981) 189.

\bibitem{Barvinsky}
A.O. Barvinsky, D. Blas, M. Herrero-Valca, S.M. Sibiryakov and C.F.
Steinwachs, {\it Renormalization of gauge theories in the
backgrorund-field approach,} JHEP {\bf 1807} (2018) 035,
arXiv:1705.03480.

\bibitem{Batalin-Lavrov}
I.A. Batalin, P.M. Lavrov and I.V. Tyutin, {\it Multiplicative
renormalization of Yang-Mills theories in the background-field
formalism,}  Eur. Phys. J. {\bf C78} (2018) 570, arXiv:1806.02552.

\bibitem{Frenkel}
J. Frenkel and J.C. Taylor, {\it Background gauge renormalization
and BRST identities,}  Annals Phys. {\bf 389} (2018) 234,
arXiv:1801.01098.

\bibitem{Batalin-Lavrov-Tyutin}
I.A. Batalin, P.M. Lavrov and I.V. Tyutin, {\it Gauge dependence and
multiplicative renormalization of Yang-Mills theory with matter
fields,}  Eur. Phys. J. {\bf C79} (2019) 628, arXiv:1902.09532.

\bibitem{LavShap} P.M. Lavrov and I.L. Shapiro,
{\it Gauge invariant renormalizability of quantum gravity,} Phys.
Rev. {\bf D100} (2019) 026018,
arXiv:1902.04687.

\bibitem{Breno}  B.L. Giacchini, P.M. Lavrov and I.L. Shapiro,
{\it Background field method and nonlinear gauges,} Phys. Lett. {\bf
B797} (2019) 134882, arXiv:1906.04767.

\bibitem{Percacci}
R. Percacci and G.P. Vacca, {\it The background scale Ward identity
in quantum gravity,} Eur. Phys. J. {\bf C77} (2017) 52,
arXiv:1611.07005.

\bibitem{Hooft}
G. 't Hooft, {\it An algorithm for poles at dimension four
in the dimensional regularization procedure,}
Nucl. Phys.  {\bf B62} (1973) 444.

\bibitem{Kluberg}
H. Klusberg-Stern and J.B. Zuber, {\it Renormalization of
non-Abelian gauge theories in a background-field gauge. I. Green's
functions,} Phys. Rev. {\bf D12} (1975) 482.

\bibitem{Grisaru}
M.T. Grisaru, P. van Nieuwenhuizen and C.C. Wu,
{\it Background field method versus normal field theory
in explicit examples: One loop divergences in S matrix and
Green's functions for Yang-Mills and gravitational fields,}
Phys. Rev.  {\bf D12} (1975) 3203.

\bibitem{Capper}
D.M. Capper and A. MacLean, {\it The background field method at two loops:
A general gauge Yang-Mills calculation,} Nucl. Phys.  {\bf B203} (1982) 413.

\bibitem{Ichinose}
S. Ichinose and M. Omote, {\it Renormalization using the background-field
formalism,} Nucl. Phys.  {\bf B203} (1982) 221.

\bibitem{Goroff}
M.H. Goroff and A. Sagnotti, {\it The ultraviolet behavior of Einstein gravity,}
Nucl. Phys.  {\bf B266} (1986) 709.

\bibitem{Ven}
A.E.M. van de Ven, {\it Two-loop quantum gravity,} Nucl. Phys.
{\bf B378} (1992) 309.

\bibitem{Grassi}
P.A. Grassi, {\it Algebraic renormalization of Yang-Mills
theory with background field method,} Nucl. Phys.  {\bf B426} (1996) 524.

\bibitem{Becchi}
C. Becchi and R. Collina, {\it Further comments on the
background field method and gauge invariant effective action,}
Nucl. Phys.  {\bf B562} (1999) 412, hep-th/9907092.

\bibitem{Ferrari}
R. Ferrari, M. Picariello and A. Quadri, {\it Algebraic aspects of
the background field method,} Annals Phys. {\bf 294} (2001) 165,
hep-th/0012090.

\bibitem{DeWitt} B.S. DeWitt,
{\it Dynamical theory of groups and fields}, (Gordon and Breach,
1965).

\bibitem{FPaction}
L.D. Faddeev and V.N. Popov, {\it Feynman diagrams for the
Yang-Mills field,} Phys. Lett. {\bf B25} (1967) 29.

\bibitem{BRS}
C. Becchi, A. Rouet and R. Stoura, {\it The abelian Higgs Kibble
Model, unitarity of the S-operator,}  Phys. Lett. {\bf B52} (1974)
344.

\bibitem{BRST}
I.V. Tyutin, {\it Gauge invariance in field theory and statistical
physics in operator formalism,}  Lebedev Inst.preprint N 39 (1975);
arXiv:0812.0580.

\bibitem{Delbourgo} R. Delbourgo and M. Ramon-Medrano,
{\it Becchi-Rouet-Stora gauge identities for gravity,}
Nucl. Phys. {\bf B110} (1976) 467.

\bibitem{Townsend} P.K. Townsend and P. van Nieuwenhuizen,
{\it BRS gauge and ghost field supersymmetry in gravity
and supergravity,}  Nucl. Phys. {\bf B120} (1977) 301.

\bibitem{JDG-QG}
J.D. Gon\c{c}alves, T. de Paula Netto and I.L. Shapiro,
{\it On the gauge and parametrization ambiguity in quantum gravity,}
Phys. Rev. {\bf D97} (2018) 026015,
arXiv:1712.03338.

\bibitem{avram-tes} I.G. Avramidi,
{\it Covariant methods for the calculation of the effective action
in quantum field theory and investigation of higher-derivative
quantum gravity,} (PhD thesis, Moscow University, 1986).
hep-th/9510140.

\bibitem{a} I. L. Shapiro and A. G. Jacksenaev,
{ \it Gauge dependence in higher derivative quantum gravity
and the conformal anomaly problem},
Phys. Lett. {\bf B324}  (1994) 284.

\bibitem{LMS} P.~Labus, T.~R.~Morris and Z.~H.~Slade,
{\it  Background independence in a background dependent
renormalization group,}
Phys. Rev.{\bf D94} (2016) 024007,
arXiv:1603.04772.

\bibitem{Alwis}  S.P.~de Alwis,
{\it Exact RG Flow Equations and Quantum Gravity,}
JHEP {\bf 1803} (2018) 118,
arXiv:1707.09298.

%
\end{thebibliography}
\end{document}